\documentclass[lettersize,journal]{IEEEtran}
\usepackage{amsmath,amsfonts}
\usepackage{algorithmic}
\usepackage{algorithm}
\usepackage{array}
\usepackage[caption=false,font=footnotesize,labelfont=footnotesize]{subfig}
\usepackage{textcomp}
\usepackage{stfloats}
\usepackage{url}
\usepackage{verbatim}
\usepackage{graphicx}
\usepackage{cite}
\usepackage{tikz}
\usepackage{multirow}
\usepackage{makecell}
\usepackage{booktabs}
\usepackage{censor}
\usepackage[all]{nowidow}
\usepackage[export]{adjustbox}

\newcommand\submittedtext{%
  \footnotesize\centering This work has been submitted to the IEEE for possible publication. Copyright may be transferred without notice, after which this version may no longer be accessible.}

\newcommand\submittednotice{%
\begin{tikzpicture}[remember picture,overlay]
\node[anchor=south,yshift=10pt] at (current page.south) {\fbox{\parbox{\dimexpr0.65\textwidth-\fboxsep-\fboxrule\relax}{\submittedtext}}};
\end{tikzpicture}%
}

\begin{document}

\title{A Grating Based High-Frequency Motion Stimulus Paradigm for Steady-State Motion Visual Evoked Potentials}

\author{Bartu Atabek,
Efecan Y\i lmaz,
Cengiz Acart\"{u}rk,
Murat Perit \c{C}ak\i r 
\thanks{
B. Atabek Author, E. Y\i lmaz Author, and M. P. \c{C}ak\i r Author are with the Middle East Technical University, Ankara, T\"{u}rkiye.}
\thanks{C. Acart\"{u}rk Author is with the Jagiellonian University, Krakow, Poland.
}}

\maketitle

\begin{abstract}
\textit{Objective: This paper proposes a novel type of stimulus in the shape of sinusoidal gratings displayed with an imperceptibly high-frequency motion. The stimulus has been designed for use in BCI (Brain Computer Interface) applications that employ visually evoked potentials (VEPs) in an effort to mitigate discomfort associated with VEPs. The stimuli set included traditional VEP stimuli, already established in the literature, allowing comparative analyses. We conducted analyses of signal distinction measures by calculating the signal-to-noise ratio and the classification performance of its evoked potentials.}
\textit{Methods: Fourteen participants were seated in a dimly lit room facing a display. Participants’ fixation on the central stimulus was controlled by means of a desktop eye tracker. Participants attended a flicker-based steady-state VEP (SSVEP) task, a motion-based steady-state-motion VEP (SSMVEP) task, and the novel stimulus task (the imperceptible grating SSMVEP). Participants were asked to complete behavioral fatigue scale tasks.}
\textit{Results: A significant effect of stimulus type was observed, accompanied by insignificant differences in prediction accuracy. Partially significant task effects were obtained in fatigue scale tasks.}
\textit{Conclusion: The study revealed that the imperceptible grating SSMVEP stimulus successfully evoked SSMVEP responses within acceptable margins in the related cortical regions. This novel stimulus contributes to BCI research by providing an imperceptible interface, improving already established stimuli design in the SSVEP and the SSMVEP literature.}
\textit{Significance: The present paper provides a novel SSMVEP stimulus type that may inform the future design of effective VEP-based BCI paradigms that allow seamless interaction with computer interfaces.}
\end{abstract}


\submittednotice
\section{Introduction}
\IEEEPARstart{T}{he} historical exploration of visual evoked potentials (VEPs) dates back to the mid-20th century, marked by the discovery of rhythmic brain waves synchronizing with a flickering light frequency \cite{adrian34}. Over time, VEPs have played a pivotal role in unraveling the complex interaction among physical stimuli, brain activity, and human cognition \cite{regan89,xie12,snowden04}. Event-related potentials (ERPs) within VEPs might further point to inter-brain connectivity with the influence of internal factors, such as activation of memory and/or motor related regions in the presence of visual stimuli.

VEPs help researchers investigate the intricate dynamics of the human brain when responding to visual stimuli, in particular with electro-physiological activations measured through electroencephalograpy (EEG) devices and its electrodes attached to the area corresponding to the occipital lobe. Among the distinct types of VEPs, steady-state VEPs (SSVEPs) have been widely used in brain-computer interface (BCI) studies. SSVEP responses are evoked by repeated visual stimuli, which are made to flicker at a particular frequency that invokes oscillatory responses at a similar frequency in the occipital cortex. SSVEPs provide advantages like simultaneous presentation of multiple stimuli and BCI based categorization mechanisms \cite{härdle14} as well as improved performance in user training.

However, the prevalent use of flicker-based stimulation in BCI studies has been associated with practical drawbacks, such as adverse effects on the human eye and fatigue responses \cite{punsawad12}. Consequently, the sensation of flickering stimuli has been linked to visual fatigue and discomfort, despite its effectiveness in invoking SSVEP stimulation with precise oscillating responses, and high pattern recognition accuracy \cite{punsawad12}. BCI research takes into account the mental workload and fatigue induced by the paradigm, as much as it prioritizes information transmission rates \cite{chai19}. Therefore, addressing these concerns is crucial for a sustainable BCI system that allows for improved usability.

A potential solution to mitigate the issues caused by fatigue in SSVEP-based BCI studies involves investigations with SSVEP that have higher stimuli flicker frequencies. Jiang et al. \cite{jiang22} highlight that SSVEP-based BCIs can adopt flickering stimuli in a wide range of frequencies with different response classification performance characteristics. While lower frequencies (\(\sim15\)Hz) invoke SSVEP responses with the most substantial amplitude \cite{pastor03}, and higher stimulation frequencies are associated with decreased SSVEP amplitude \cite{bakardjian10,hoffman09,regan77}. The critical fusion frequency (CFF) marks a threshold beyond which visual stimuli elicit diminished cortical activity \cite{williams04}. On the other hand, despite this reduced activity, studies on high-frequency SSVEP BCI have shown that they offer lower levels of discomfort and safety for prolonged BCI use. A consequence is that these higher frequencies fall beyond the critical range for epileptic photosensitive individuals \cite{ladouce22}. Further research claimed that visual stimuli with frequencies above 50Hz are considered safer for individuals with photosensitive epilepsy, in contrast to the most provocative frequency of 15-25Hz for inducing seizures \cite{zhu10,fisher05}. They suggest that imperceptible flickering stimuli with frequencies above the CFF can reduce fatigue, especially when BCIs are used for daily communication or assisted movement \cite{chen19,maye17}, which define prolonged periods of interaction.

In an investigation of high-frequency stimuli, Ladouce et al. \cite{ladouce22} indicated that while this type of invoking SSVEP responses may reduce overall visual discomfort, their classification performance may not be competitive enough to design a reliable and responsive BCI. Furthermore, high-frequency stimuli also require extended stimulation periods to achieve optimal performance \cite{ladouce22} in classification of invoked responses from the cortical regions. For example, Chen et al. investigated this distinction between low and high-frequency SSVEP-based BCI by employing a 45-target spelling system. Their results showed that high-frequency stimulation achieved higher accuracy but at the cost of longer data collection times \cite{chen14}.

In response to these challenges posed by flicker-based SSVEPs, a recent breakthrough emerged in the form of steady-state motion visual evoked potentials (SSMVEPs). Xie et al. \cite{xie12} introduced the motion paradigm, targeting SSMVEPs, and replacing flicker stimulation with motion, to reduce visual discomfort and the resulting fatigue effects. This paradigm, rooted in the two-streams hypothesis of the visual system \cite{duszyk14}, exploits the ventral stream's sensitivity to object characteristics (the \emph{what} pathway), and to object speed and direction of movement (the \emph{where} pathway). By simultaneously activating both pathways, motion stimulation introduces a more robust brain response compared to flicker-based SSVEP stimuli, which only modulates the luminance of objects \cite{xie12}.

Furthermore, Chai et al. \cite{chai20} provided empirical support for the effectiveness of SSMVEP stimuli based on non-directional motion in reducing visual fatigue. Their findings demonstrate that this approach induces visual-related potentials in the brain without evident adaptation phenomena. This promising avenue not only enhances classification accuracy but also addresses issues of visual fatigue, making it a compelling alternative for BCIs.

In the present study, we propose a novel stimulus design that allows future prospects for expansion toward imperceptible SSMVEP studies. The design targets reduced levels of discomfort and computationally tractable stimuli categorization in real-time, using high-frequency visual stimuli. To this end, we utilized a Gabor patch, which is a sinusoidal grating, with predefined features, for a comparative analysis against previous stimuli design in the relevant SSVEP and SSMVEP literature \cite{yan17,yan18,norcia16}. The Gabor patch design allowed a better level of control in real-life applications, compared to traditional stimuli that require high-frequency and high contrast, such as the checkerboard \cite{ne23} or other motion-based designs \cite{yan18,norcia16}.

We utilized eye tracking with a state-of-the-art desktop eye tracker, SR-Research EyeLink 1000 Plus, in order to provide a blind control for ensuring the gaze fixations of the participants on the target stimulus. We also employed a frequently used fatigue inventory, VAS-F \cite{shahid11}, to reveal the perceived fatigue of the participants, as a measure of subjective, conscious report of user experience. 

The design of the study includes three within-subject experimental tasks: An SSVEP analogous task, an SSMVEP analogous task (as per \cite{yan17, yan18, norcia16}), and the Gabor grating stimulus designed for the purpose of the study. In the Gabor design, we applied an imperceptible, high-frequency motion (SSMVEP design) so that the stimulus seemed static to the naked eye.

\section{Methodology}

\subsection{Participants \& Ethical Statement}

The experiment protocol was approved by the Middle East Technical University Human Subjects Research Ethics Committee (protocol no. 0458-ODTUIAEK-2023). Six males and eight females, aged between 20 and 44 years (M = 30.4, SD = 7.4), were recruited as participants. The participants were physically healthy at the time of recording, and all had reportedly normal or corrected-to-normal vision, with no prior diagnosis of photosensitive epilepsy. A written informed consent was obtained from all participants prior to the experiment. 

\subsection{Experiment Setup}
\begin{figure}[!htbp]
  \centering
  
  \subfloat[]{\includegraphics[width=.55\linewidth]{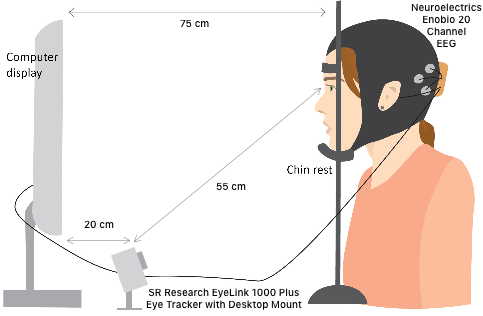}%
    \label{experimental-setup}}\hfill
  \subfloat[]{\includegraphics[width=.35\linewidth,height=3cm]{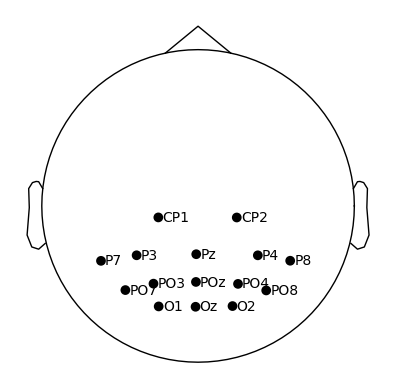}%
    \label{electrode-placement}}

    \caption{(a) Experimental setup. (b) EEG electrode placement according to the international 10-20 system.}
\end{figure}

Each participant was seated in a dimly lit room (0.83 lux measured by the chin rest during stimuli offset and/or at an empty black screen) in front of a monitor and placed their chins on a chin rest to ensure stable eye tracking accuracy. The chin rest was located 75cm away from the monitor, as shown in Figure \ref{experimental-setup}. In the experiment setting, we employed a 27" W-LED backlit, flicker-free LCD monitor with a 144Hz vertical refresh rate. The active area to render the stimuli was adjusted to a 24" display in the field of vision to match SR Research recommendations for optimal display size of degrees covered by the monitor in the visual field. Hence, we defined an active viewing rectangle corresponding to a 24" monitor with an aspect ratio of 16:9 with 1920 x 1080 pixels resolution.

\vspace{-0.5cm}
\subsection{Equipment}

\subsubsection{Eye Tracker}
Eye gaze was monitored using an SR Research EyeLink 1000 Plus eye tracker with desktop mount \cite{sr23} configured with 1000Hz sampling rate, and the eye-tracker was located in accordance with the manufacturer's instructions as shown in Figure \ref{experimental-setup}. SR-Research recommends an optimal screen size of up to 24". As we used a larger monitor, we moved the display and the tracker accordingly to match the size of a 24" display in the field of vision.

\subsubsection{EEG equipment}
EEG data was acquired using an Enobio system from Neuroelectrics (hereafter, NE) \cite{ne23} with 14 dry (Ag-CL) electrodes located in the parietal and occipital regions using the 10-20 distribution (PO7, O1, PO3, Pz, Oz, PO4, O2, PO8, CP1, CP2, P3, P7, P4, P8) at a sampling rate of 500Hz with electrode positions shown in Figure \ref{electrode-placement}. An additional virtual electrode POz was generated during the offline analysis by averaging the electrodes Pz and Oz due to the limitations of the standard EEG caps from NE. For all participants, both the reference and ground electrodes were placed on the right earlobe.

\subsubsection{Environment Sensor}
An Arduino Uno R3-based ambient light meter was built using a BH1750 light sensor module in order to measure the total illuminance by the chin rest in lux units during the course of the experiment.

\subsection{Stimuli Design}
\begin{figure}[!h]
    \centering
    
    \subfloat[]{\includegraphics[width=0.2\linewidth]{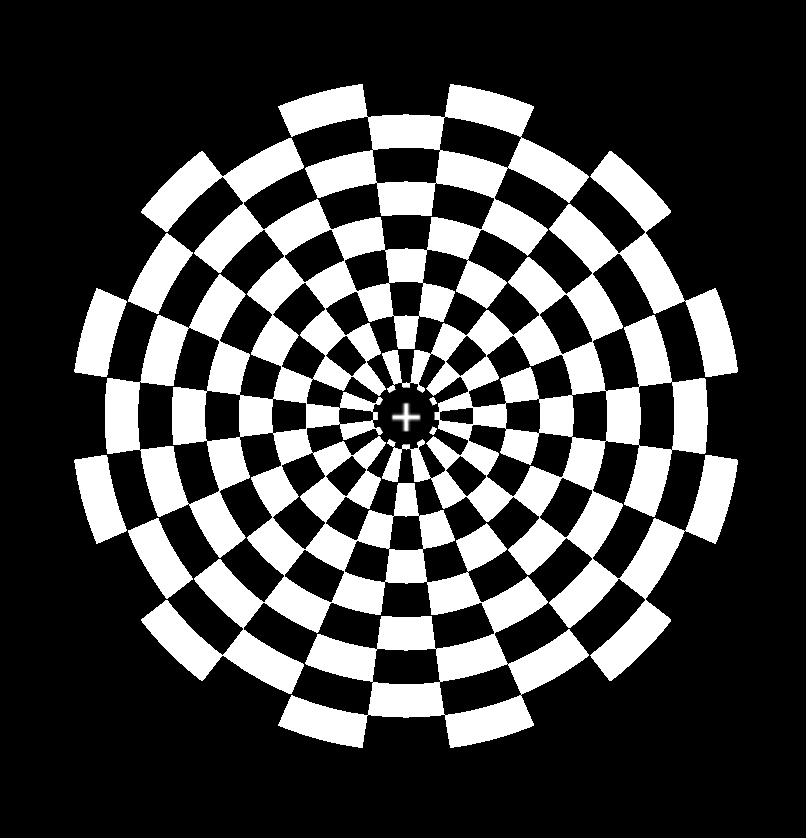}%
    \label{fig:checkerboard}}
    \hfill
    \subfloat[]{\includegraphics[height=1.83cm]{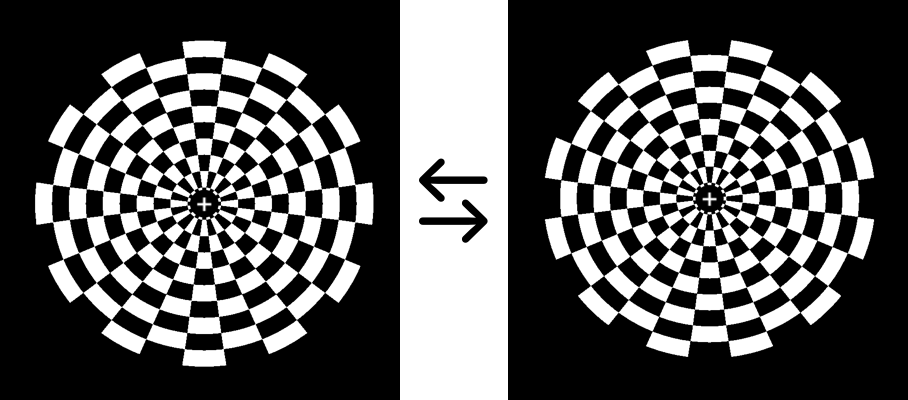}%
    \label{fig:checkerboard-reversal}}
    \hfill
    \subfloat[]{\includegraphics[width=0.2\linewidth, height=1.83cm]{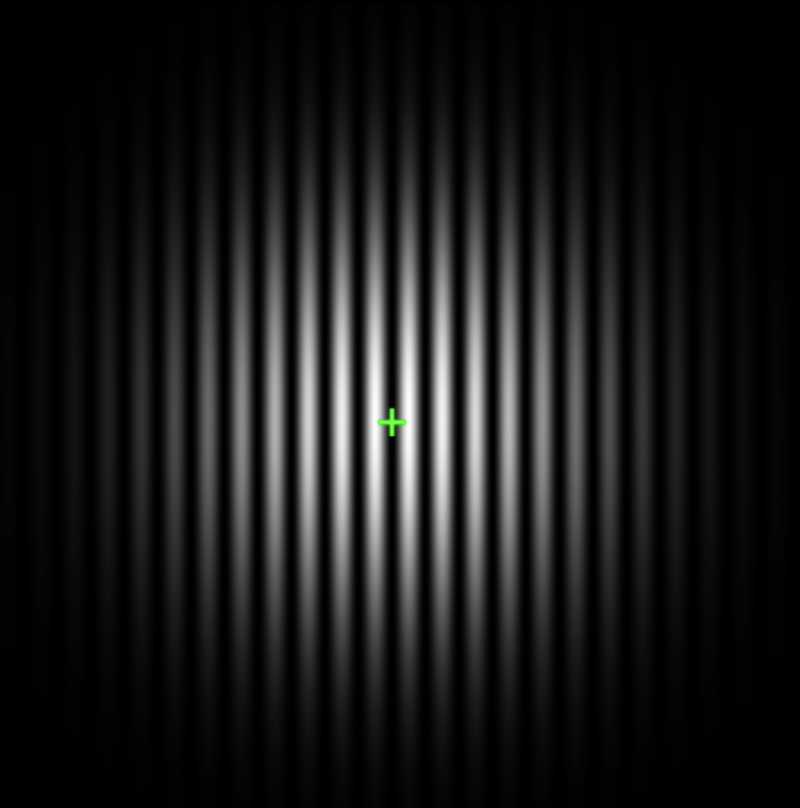}%
    \label{fig:grating}}
    
    \caption{(a) The checkerboard stimulus used through tasks 1 and 2. (b) is the two contrasting stimulus alternated at a specified frequency for pattern reversal SSVEP. (c) is the grating stimulus used for the imperceptible high-frequency SSMVEP.}
    \label{fig:stimulus-types}
\end{figure}

PsychoPy and PsychoPy Vision Science libraries \cite{psychopy} were used to design and implement three different stimulus paradigms featuring pattern reversal (checkerboard SSVEP reference condition, Figure \ref{fig:checkerboard}), radial contraction expansion (checkerboard SSMVEP reference condition, Figure \ref{fig:checkerboard-reversal}), and imperceptible pulsation display methods (Figure \ref{fig:grating}).

\vspace{-.5cm}

For the initial two tasks, the same black and white checkerboard stimulus was displayed, comprising five radial cycles, each containing 12 angular cycles. At the center, a white circle with a radius of 25 pixels and a fixation cross were positioned to help participants fixate at the center during stimuli displays. The checkerboard stimuli are defined by several parameters, including the subtended visual angle of each tile (spatial frequency), the number of reversals per second, mean luminance, field size, and pattern contrast.

For the third task, the newly proposed Gabor grating stimulus was displayed. Gabor patches have a set of defining properties, including contrast, phase, spatial frequency, and amplitude. Similarly, with the first two tasks, we placed a fixation cross at the center of the Gabor patch to ensure a fixed salient spot was provided for all participants.

\subsubsection{Pattern Reversal Checkerboard Stimuli for SSVEP}
Pattern reversal stimuli involving the oscillatory alternation of graphical patterns on a computer screen were employed. These stimuli consisted of at least two patterns alternated at a specified rate of alternations per second \cite{odom04}, which define the display.

\subsubsection{Radial Contraction-Expansion Motion Checkerboard Stimuli for SSMVEP}
The contraction-expansion motion, based on a sine wave as designed by Yan et al. in 2016 \cite{ne23}, was utilized. Visual stimulation was presented to the participants on a computer screen, with a screen refresh rate of \( f_r \). Frame images were generated using Formula (1):

\begin{equation}
\phi(t)=\frac{\pi}{2}+\frac{\pi}{2}\cdot sin\left(2\pi\cdot\mathrm{f}_{c}\cdot t - \frac{\pi}{2}\right)
\end{equation}

\noindent
where \( f_c \) is the motion frequency, representing the reciprocal of the required time for one contraction-expansion cycle of the checkerboard. The checkerboard contracts when the phase \( \phi(t) \) is turned from 0 to \( \pi \), and it expands when \( \phi(t) \) is turned from \( \pi \) to 0. The screen refresh rate is designated as \( fr \), and \( n = 1, 2, 3, \ldots \) represents the frame number.

\subsubsection{High-Frequency Grating Based Motion for SSMVEP}
The motion animation to evoke an SSMVEP response was created using a sinusoidal function with Gaussian masking around the Gabor patch. The Gabor patch itself was characterized by a phase value of 2.5 and a spatial frequency value of 25Hz. The Gaussian mask surrounding the stimulus underwent pulsing motion at high frequencies beyond the CFF, specifically at 72Hz. This motion was validated using high-speed cameras, as it was imperceptible to the naked eye.

\subsection{Data Acquisition}
The stimulus presentation program was an automated and customizable software application developed in Python using PsychoPy version 2023.2.3 \cite{psychopy}. It generates the visual stimuli and builds the experiment flow. We used the EyeLink Developers Kit \cite{eyelink_sdk} for eye tracking and the Neuroelectrics Instrument Controller (NIC2) \cite{nic2} to record and stream the EEG data. Experiment markers for stimuli onset and dilation baseline markers, eye tracking data, EEG data, and environment data streams were synchronized using Lab Streaming Layer (LSL) \cite{lsl}.

Each participant had their gaze calibrated using EyeLink's calibration routine available in the SDK. For best results, we utilized 9 calibration points spanning across the active tracking area. After calibration, an automatic calibration validation, again through EyeLink SDK, was performed to ensure an accurate calibration. We repeated the full calibration procedure if accuracy and precision RMS of either right or left eye were above 1.5$^\circ$ and 0.5$^\circ$ respectively.

\subsection{Experimental Protocol}

\begin{figure}[!h]
\centering

\subfloat[The overall experimental process.]{\includegraphics[width=\linewidth, height=1.5cm]{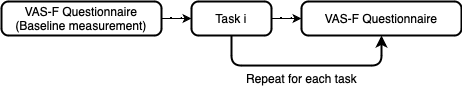}%
\label{fig:protocol_1}}

\subfloat[Contents of each task.]{\includegraphics[width=\linewidth]{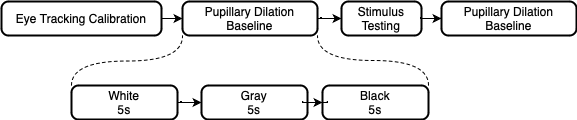}%
\label{fig:protocol_2}}

\caption{The experimental flow detailing the blocks and contents.}
\label{fig:experimental_flow}
\end{figure}

The experiment was divided into three tasks each containing 4 central blocks: calibration, dilation baseline, stimulus testing (experiment session), and visual analogue scale to evaluate fatigue (VAS-F) severity questionnaire. The experimental process is shown in Figures \ref{fig:protocol_1} and \ref{fig:protocol_2}.

\subsubsection{Eye Tracking Calibration and Baseline Procedures}
In order to keep a record of the quality of the EEG recording, baseline neuro-electric activity was monitored for 10 seconds while the participant rested their chins on the chin-rest. This was followed by the calibration of the eye-tracker as explained in the \emph{Data Acquisition} section above.

\subsubsection{Experiment}
The experiment sessions consisted of three tasks for testing each of the three stimulus types. During the task, EEG and gaze data were simultaneously recorded while participants gazed at the specified target stimuli. The tasks were presented in random order for each participant to mitigate task order and fatigue effects.

\begin{figure}[!htbp]
\centering

\subfloat[The experiment process of task 1.]{\includegraphics[width=\linewidth]{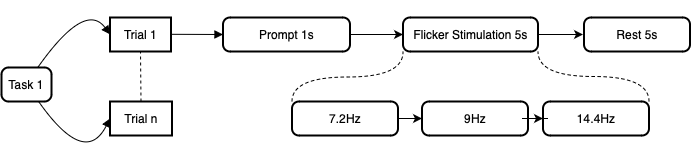}%
\label{fig:protocol_3}}

\subfloat[The experiment process of task 2.]{\includegraphics[width=\linewidth]{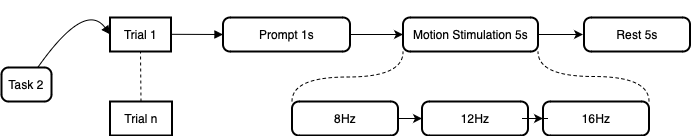}%
\label{fig:protocol_4}}

\subfloat[The experiment process of task 3.]{\includegraphics[width=\linewidth]{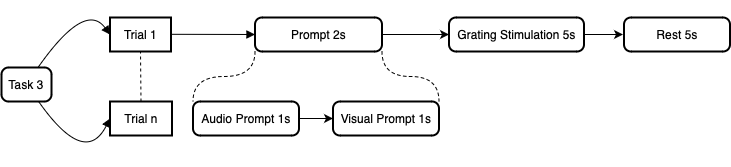}%
\label{fig:protocol_5}}

\caption{The design of the experiment sessions.}
\end{figure}

\emph{Task 1 - Black-white Pattern Reversal Checkerboard Stimulation:} Three types of flickering checkerboard stimulations were displayed on the screen side by side. The frequencies were set at 7.2Hz, 9Hz, and 14Hz, respectively. During each trial, the participants were requested to look at the target stimulus marked with a fixation cross at the center for 5 seconds. The task included 30 trials in total, which were interleaved with a resting episode of 5 seconds.

\emph{Task 2 - Radial Contraction-expansion Motion Checkerboard Stimulation:} The experimental contents and the task flow were the same as Task 1, except the motion frequency \( f_c \)  was set at 8Hz, 12Hz, and 16Hz, respectively, for each target. 

\emph{Task 3 - High-frequency Grating-based Motion Stimulation:} A single stimulus with a motion frequency of 72Hz was displayed on the screen. Stimulation lasted 5 seconds per trial, 30 trials in total, with a rest interval of 5 seconds. In this task, the participants were asked to close their eyes and reopen them after they heard an auditory signal during the rest interval. This method is used so that the CCA algorithm can be applied to distinguish between offset and onset cases of the stimuli. This was a technical challenge as the present form of the imperceptible motion animation is at an early development stage, which currently necessitates the entire screen to be refreshed. Therefore, only a single frequency could be displayed at a time in this study, in contrast to the typical SSVEP and SSMVEP tasks in the literature featuring simultaneous presentation of multiple stimuli with different flicker/motion frequencies.

The experiments were designed to compare the differences in brain responses evoked by the flickering black-white checkerboard, the contracting-expanding checkerboard, and the grating stimuli with high-frequency pulsing motion to determine the feasibility of improving BCI usability by using Gabor patches with imperceptible motion stimulation. The first two conditions mainly served as a validation of our experiment setup and data processing pipeline, as well as a basis for contrasting the level of evoked responses by the newly proposed imperceptible motion stimulation VEP paradigm to the well-known perceptible SSVEP and SSMVEP paradigms.

\subsection{Fatigue Questionnaire}
Upon completing each task, the participants were asked to fill out the Visual Analogue Scale to Evaluate Fatigue Severity (VAS-F) questionnaire \cite{shahid11} to self-report their perceived levels of fatigue and energy. Participants also filled out the questionnaire at the beginning of the experiment to establish a baseline for task-induced fatigue level comparisons.

\section{Data Processing}
After each session of data recording was complete, the stored LSL \emph{xdf} files were parsed into \emph{csv} format and analyzed offline using the MNE \cite{gramfort13} library in Python.

\subsection{Pre-Processing of the EEG Data}
Following the start and end times of each trial, the stimulation data segments were extracted. Then, a band-pass filter based on the perimeters of the target frequencies (e.g., for task 1, it was from 7Hz to 15Hz) was imposed to attenuate high-frequency interference and low-frequency drifts. The ZapLine method \cite{cheveigné19} was employed to mitigate the impact of power line artifacts in our EEG recordings. Finally, the Artifact Subspace Reconstruction (ASR) algorithm \cite{mullen15} was utilized to attenuate the effects of eye movement and muscle artifacts on the EEG data.

\subsection{Power Spectral Density}
The frequency spectrum of the EEG data was analyzed via Fast Fourier Transform (FFT) \cite{bach99}. We applied plain FFT as a special case of Welch’s method in the MNE toolbox, with a single Welch window spanning the entire trial and no specific windowing function (i.e., applying a boxcar window). The first second following the onset of each trial was excluded from FFT analysis since the response of steady-state often takes a while to stabilize, and the transient phase, in the beginning, can distort the signal estimates.

\subsection{Signal-to-Noise Ratio}
In reactive BCI applications, signal-to-noise ratio (SNR) is considered an important indicator for assessing the effectiveness of stimulation paradigms such as SSVEP or SSMVEP \cite{meigen99}. In the analyses, two critical parameters were needed for SNR spectrum computation: the number of noise bins and whether to skip bins directly adjacent to the target bin. Adjusting these parameters significantly impacts the spectrum. We choose to skip one bin adjacent to the target, comparing the power at each bin with the average power of three neighboring bins on each side. This approach balances sensitivity to power modulations while minimizing artifacts from extreme parameter choices.

\subsection{Filter Bank Canonical Correlation Analysis}
Canonical correlation analysis (CCA) is a widely utilized analysis method for target recognition in both SSVEP and SSMVEP paradigms. Filter Bank Canonical Correlation Analysis (FB-CCA) is an extension of Canonical Correlation Analysis (CCA) that incorporates filter banks to analyze multi-channel signals. In this study, all EEG sampling channels were selected as a set of variables; then, filter banks were employed to decompose them into different frequency bands. This decomposition facilitates the identification of correlated patterns in specific frequency ranges, which are used to calculate the canonical correlation coefficients with the generated reference signals. The maximum target of the correlation coefficient is considered to be the focused target \cite{chen15}.

\subsection{Statistical Analysis}
The statistical evaluation of the present experiment setting involved four distinct analyses. Firstly, a descriptive analysis of the power spectrum corresponding to the trials with the imperceptible Gabor grating stimulus was conducted to observe whether there was a detectable peak power increase around the target frequency range of 72Hz. Secondly, we compared the SNR levels obtained during trials with the imperceptible Gabor grating stimulus with the SNR values for the visible SSVEP and SSMVEP conditions. Thirdly, we contrasted the maximum CCA correlation values obtained during the SSVEP, SSMVEP, and the imperceptible Gabor grating stimulus conditions. CCA was utilized in an exploratory and controlling manner in post-processing to help us understand how an online categorization would have performed on the three tasks based on the peak canonical correlation values detected for each stimulus type. Finally, we contrasted the three conditions in terms of the self-assessed fatigue and energy levels reported by the participants via the VAS-F questionnaire. The JASP software \cite{jasp} was utilized for the statistical analysis.

\section{Results}
\subsection{Frequency Analysis Results}
\begin{figure}[!h]
  \centering
  
  \subfloat[Subject 1]{\includegraphics[width=.5\linewidth]{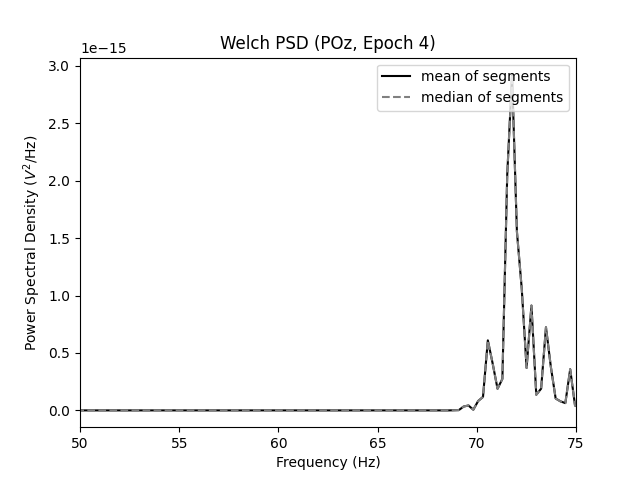}%
    \label{fig:sub1}}\hfill
  \subfloat[Subject 4]{\includegraphics[width=.5\linewidth]{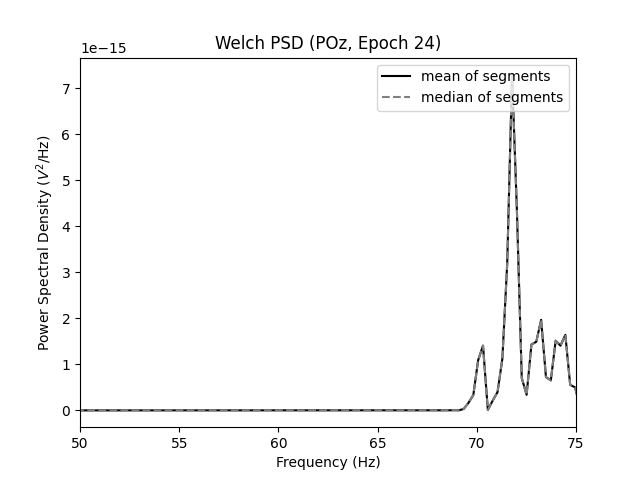}%
    \label{fig:sub4}}\hfill
    
  \subfloat[Subject 5]{\includegraphics[width=.5\linewidth]{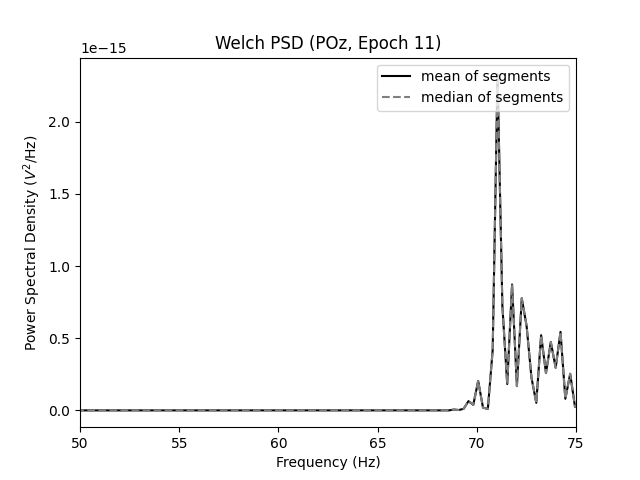}%
    \label{fig:sub5}}\hfill
  \subfloat[Subject 6]{\includegraphics[width=.5\linewidth]{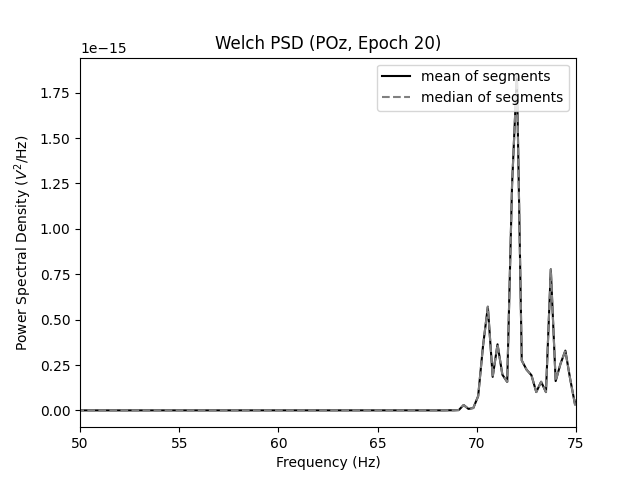}%
    \label{fig:sub6}}\hfill
    
  \subfloat[Subject 7]{\includegraphics[width=.5\linewidth]{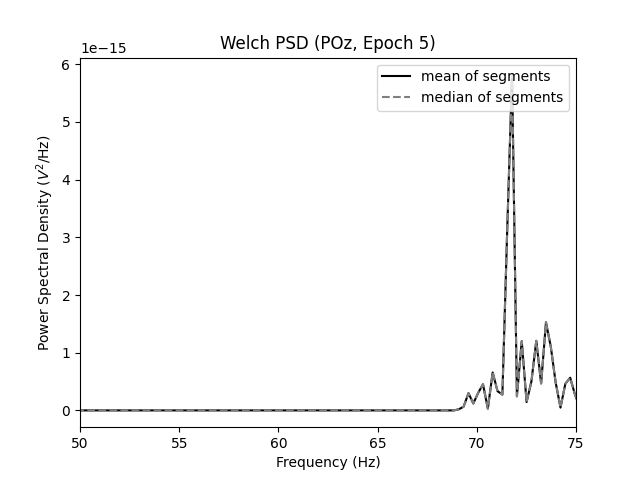}%
    \label{fig:sub7}}\hfill
  \subfloat[Subject 9]{\includegraphics[width=.5\linewidth]{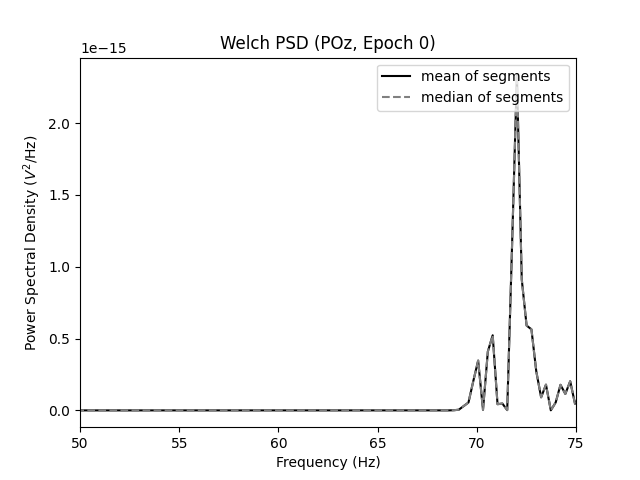}%
    \label{fig:sub9}}\hfill

  \subfloat[Subject 11]{\includegraphics[width=.5\linewidth]{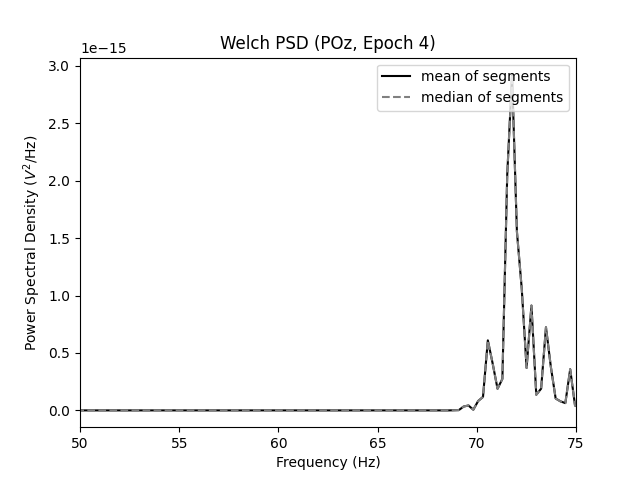}%
    \label{fig:sub11}}\hfill
  \subfloat[Subject 13]{\includegraphics[width=.5\linewidth]{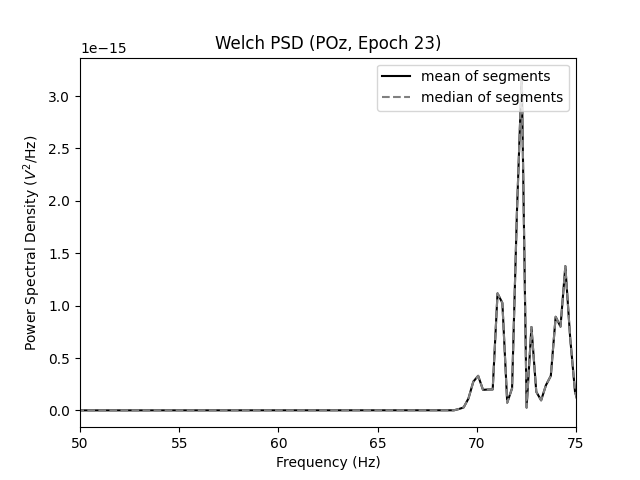}%
    \label{fig:sub13}}\hfill
  
  \caption{Frequency domain analysis for SSMVEPs induced by the imperceptible grating based pulsing motion paradigm for a selected sample of subjects.}
  \label{fig:psd}
\end{figure}

The Welch power spectrum \cite{barb10} analysis results for a sample of participants for the imperceptible Gabor grating stimulus is shown in Figure \ref{fig:psd}. It can be seen in the power spectrum plots that there are distinguishable peak amplitudes around the motion frequency of 72Hz for all subjects. 

\vspace{-0.75cm}

\subsection{SNR Analysis Results}
\begin{figure}[!htbp]
  \centering
  \includegraphics[width=\linewidth]{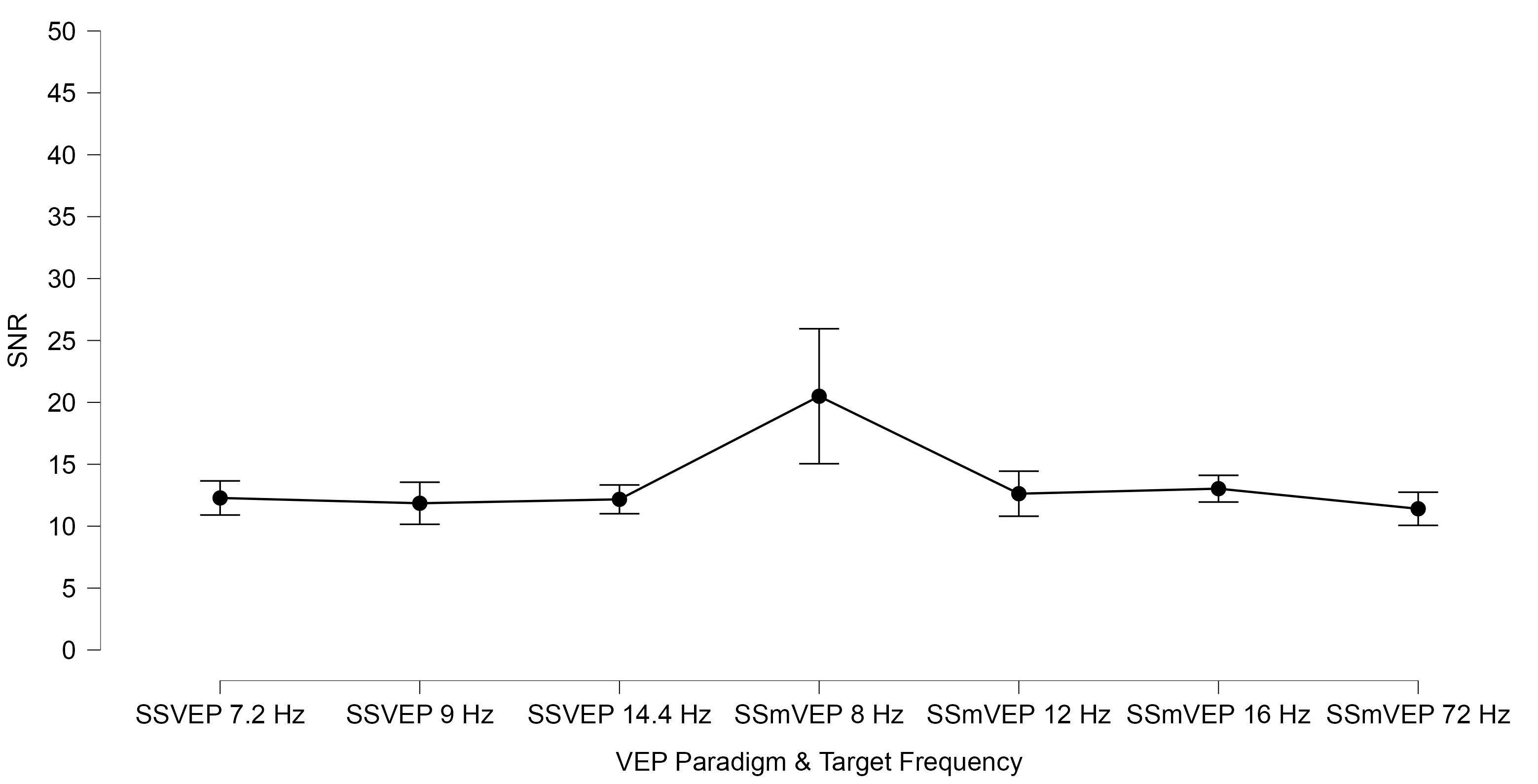}
  \caption{The average SNR values for each of the used VEP paradigm and target frequency.}
  \label{fig:snr_avg}
\end{figure}

\begin{figure}[!htbp]
  \centering
  \includegraphics[width=\linewidth]{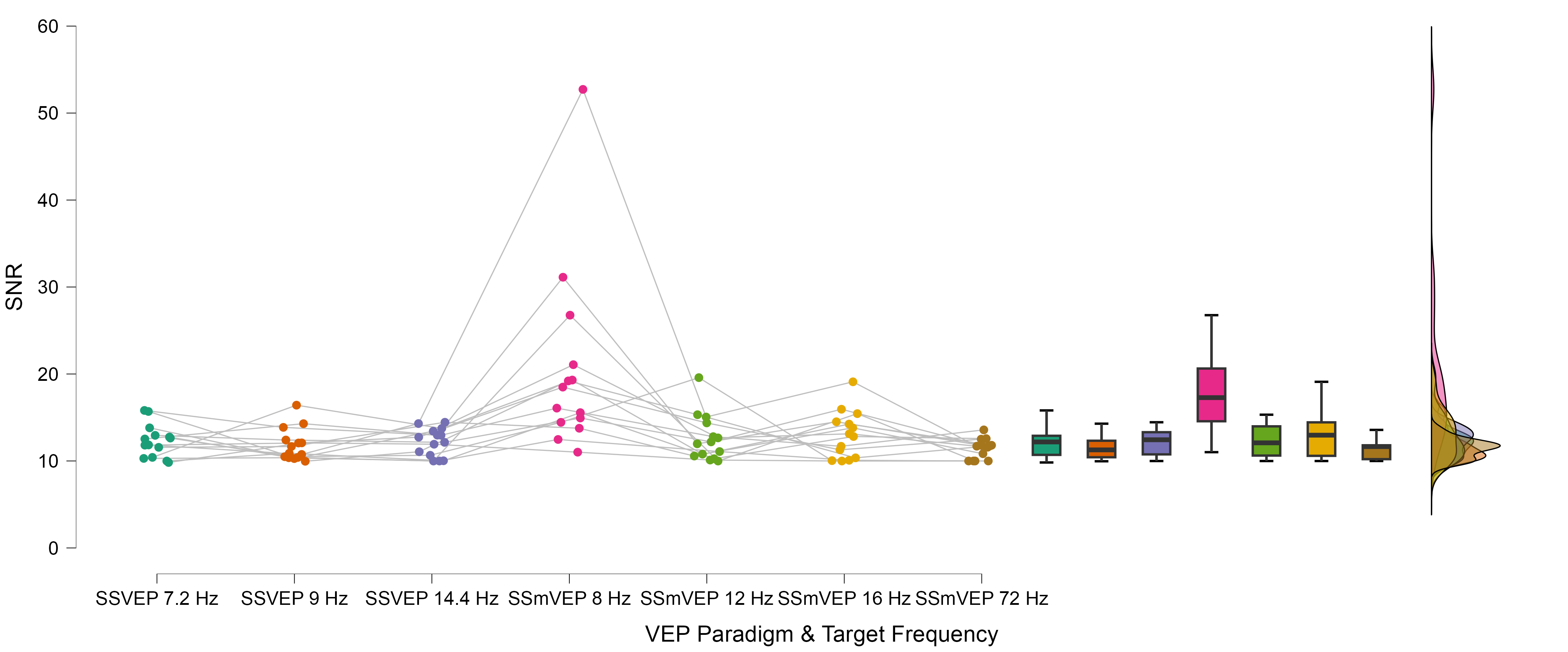}
  \caption{The distribution of SNR values for each participant for each of the used VEP paradigm and target frequency.}
  \label{fig:snr_subjects}
\end{figure}

Figure \ref{fig:snr_avg} shows the average SNR values observed for each VEP stimulus condition, and Figure \ref{fig:snr_subjects} shows the underlying distribution of those SNR values across subjects and conditions. The SNR values observed for the novel stimulus condition at 72Hz ranged between 9.99 and 13.57. A repeated measures ANOVA analysis was conducted to compare the mean SNR values obtained for each VEP stimulus type at the fundamental frequency. A significant effect of stimulus type was observed, $F(6, 78)=3.29$, $p<.01$, $\eta^2=.38$. Follow-up post-hoc tests with Holm correction showed that the effect is due to the difference between the SNR values obtained in the 8Hz SSMVEP condition compared to the other stimulus conditions. In particular, three participants registered atypically high SNR values in the 8Hz condition. However, it is known that the evoked responses might significantly differ among individuals at the presented stimuli frequency range \cite{jukiewicz18}. The remaining pairwise comparisons did not reveal any significant differences, suggesting that the imperceptible SSMVEP stimulus elicited comparable SNR levels compared to frequently used perceptible SSVEP and SSMVEP stimulus types. These findings indicate that the imperceptible grating-based pulsing motion paradigm can be employed for practical reactive-BCI applications with dry EEG setups.

\subsection{Classification Accuracy Results}
\begin{figure}[!htbp]
  \centering
  \includegraphics[width=\linewidth]{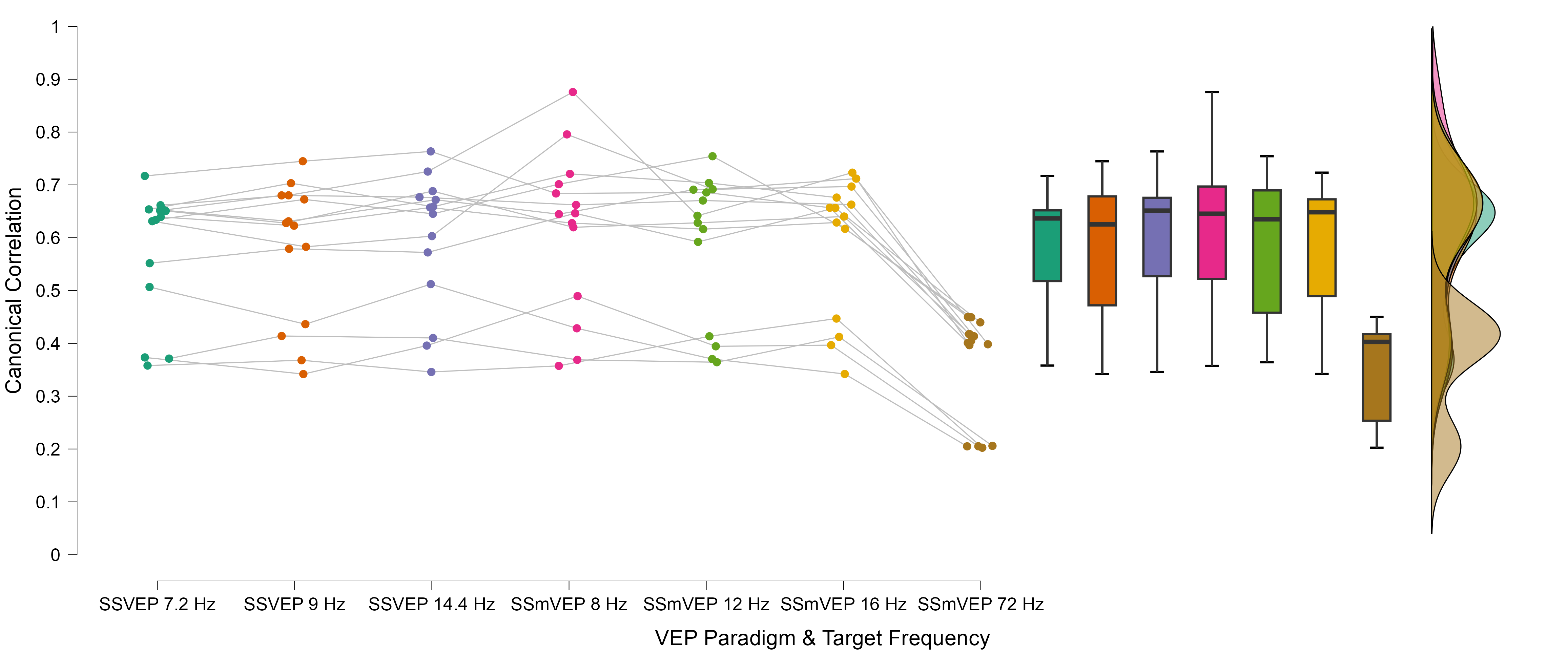}
  \caption{The distribution of correlation values for each participant for each of the used VEP paradigm and target frequency.}
  \label{fig:cca_subjects}
\end{figure}

\begin{figure}[!htbp]
  \centering
  \includegraphics[width=\linewidth]{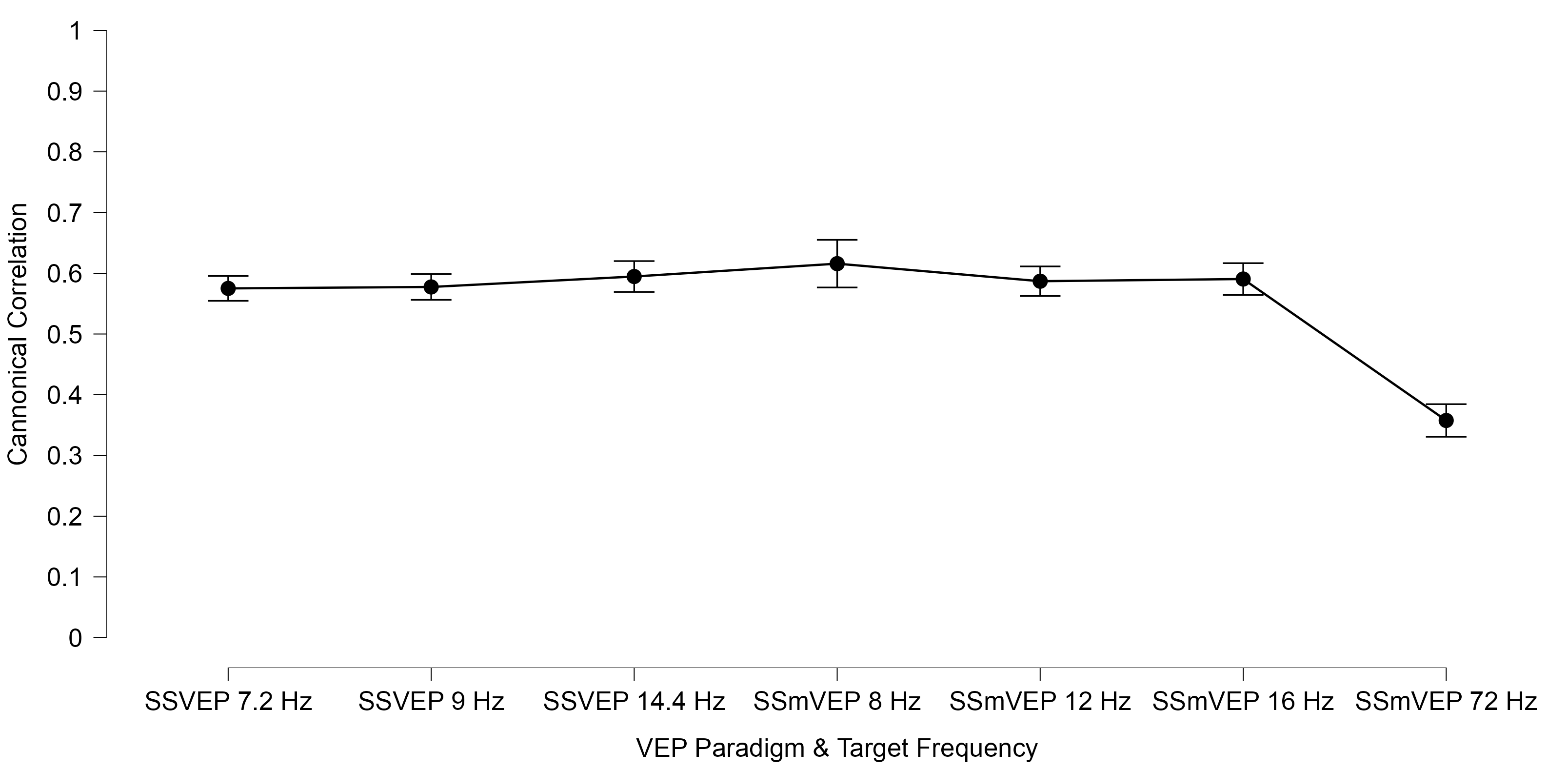}
  \caption{The average correlation values for each of the used VEP paradigm and target frequency.}
  \label{fig:cca_avg}
\end{figure}

CCA analysis indicated that the invisible SSMVEP condition elicited a mean correlation value of 0.358, which is lower than the correlation values obtained for the visible SSVEP and SSMVEP stimuli displayed in Figure \ref{fig:cca_avg}. When contrasted with perceptible stimulus types through a repeated measures ANOVA, the correlation levels obtained in the imperceptible SSMVEP condition were significantly lower, $F(6, 78) = 51.11$, $p < .01$, $\eta^2 = .78$, which is expected for stimuli in the invisible range \cite{ladouce22}. A closer analysis of each participant’s responses, as seen in Figure \ref{fig:cca_subjects} revealed that 4 participants who also had lower correlation values for the perceptible flickerıng and motion conditions have registered considerably low CCA values, whereas other participants’ correlation values were closer to the measures obtained with perceptible stimuli types. While this outcome aligns with anticipated results, a median CCA value of 0.4 elicited by the imperceptible SSMVEP stimulus is promising for the effective detection of reactive BCI applications.

Since the entire screen has to be refreshed to elicit the imperceptible SSMVEP effect, we could not present multiple imperceptible motion stimuli moving at different frequencies on the same display. Thus, we could not perform a classification accuracy analysis based on CCA results. We conducted accuracy analysis for the perceptible SSVEP and SSMVEP conditions only to validate our EEG recording setup with dry electrodes. 

\begin{figure}[!hb]
  \centering
  \includegraphics[width=0.5\linewidth]{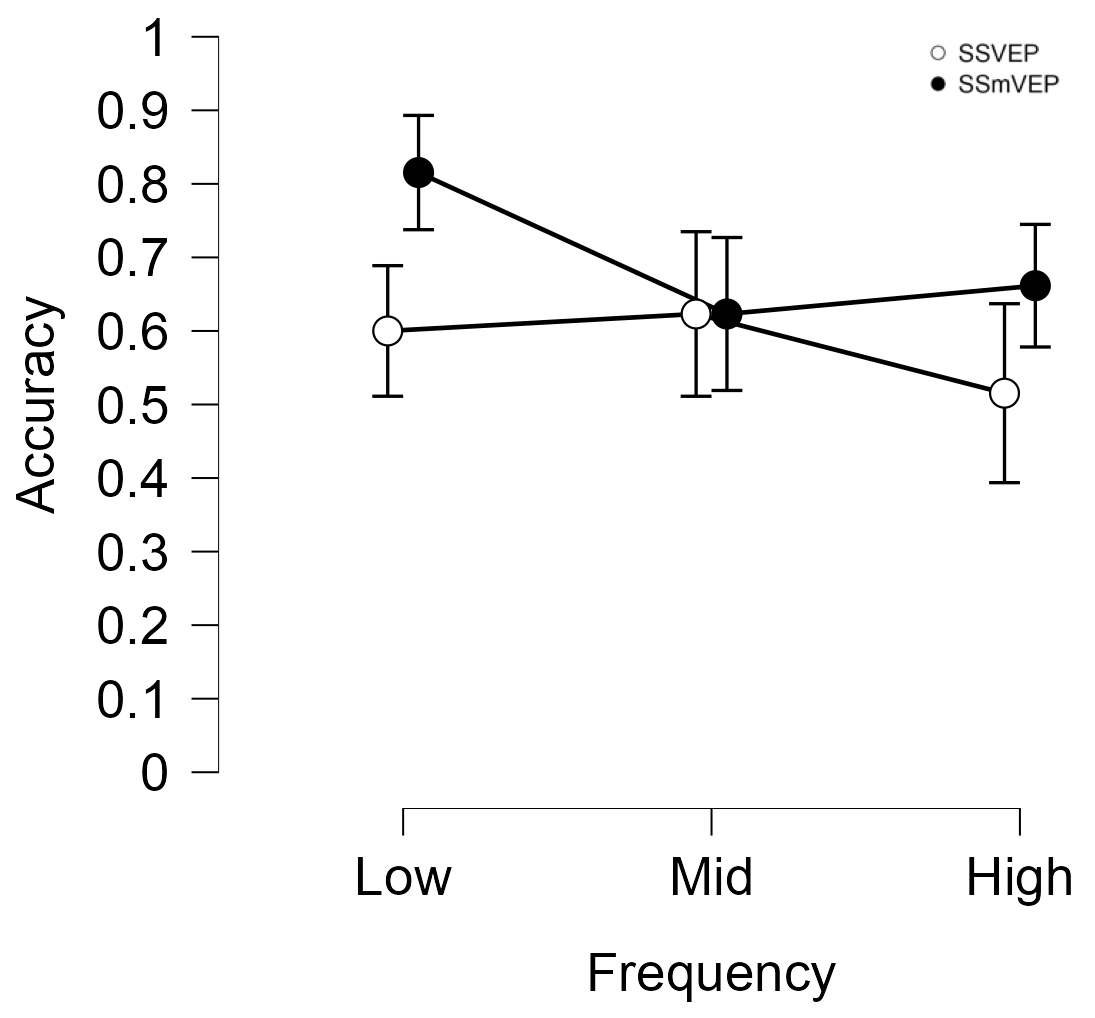}
  \caption{Average accuracy values observed for the visible SSVEP and SSMVEP conditions.}
  \label{fig:cca_visible}
\end{figure}

Figure \ref{fig:cca_visible} shows the average accuracy values obtained for each visible SSVEP and SSMVEP condition in our experiment. The average accuracy ranged between 0.52 and 0.82, which is above the by-chance level of 0.33 for all conditions. We detected significantly higher classification accuracy in the visible SSMVEP condition (M=0.70, SE=0.05) as compared to the visible SSVEP condition (M=0.58, SE=0.05), $F(1, 24) = 9.39$, $p < .05$, $\eta^2 = .13$, which was due to the high accuracy obtained for the visible SSMVEP condition at 8Hz. These results are compatible with previous studies comparing visible SSMVEP and SSVEP stimuli \cite{yan18}, which indicates that our setup succeeded in replicating the reported results in the literature.

\begin{table*}[!htbp]
  \centering
  \caption{Performance and Fatigue Levels for Each Participant and Task}
  \label{tab:performance}
  \begin{tabular}{lccc ccc ccc}
    \hline\hline
    \addlinespace[1ex]
    \multicolumn{1}{c}{\multirow{2}{*}{Subject}} & \multicolumn{3}{c}{Task 1} & \multicolumn{3}{c}{Task 2} & \multicolumn{3}{c}{Task 3} \\
    \cmidrule(lr){2-4} \cmidrule(lr){5-7} \cmidrule(lr){8-10}
    & SNR\textsuperscript{a} & Accuracy & Fatigue\textsuperscript{b} & SNR\textsuperscript{a} & Accuracy & Fatigue\textsuperscript{b} & SNR\textsuperscript{a} & Accuracy & Fatigue\textsuperscript{b} \\
    \addlinespace[1ex]
    \hline
    \addlinespace[1ex]
    S1 & 12.2059 & 53.33\% & 0.6154 & 14.7166 & 83.33\% & 0.4615 & 12.5069 & 100\% & 0.7692 \\
    
    S2 & 13.5227 & 60.00\% & -0.2308 & 14.6323 & 90.00\% & -0.9231 & 10.8347 & 100\% & 0.6923 \\
    
    S3 & 13.2573 & 70.00\% & 0.6923 & 14.6323 & 76.66\% & 1.2308 & 13.5726 & 100\% & 0.2308 \\
    
    S4 & 10.3004 & 16.66\% & 0.6154 & 12.8752 & 30.00\% & -0.2308 & 12.5664 & 100\% & 0.4615 \\
    
    S5 & 12.5012 & 70.00\% & 0.4615 & 15.6141 & 83.33\% & 1.3846 & 11.5796 & 100\% & 1.8462 \\
    
    S6 & 14.2135 & 33.33\% & 3.6923 & 19.2016 & 60.00\% & 0.4615 & 11.7851 & 100\% & 1.0769 \\
    
    S7 & 12.5600 & 30.00\% & 1.7692 & 12.8561 & 50.00\% & 1.7692 & 11.8052 & 100\% & 2.0769 \\
    
    S8 & 12.2398 & 60.00\% & 1.0000 & 14.9605 & 86.66\% & -1.0769 & 11.7349 & 100\% & 1.6154 \\
    
    S9 & 12.4972 & 70.00\% & -0.0769 & 14.3607 & 56.66\% & 0.0000 & 11.6035 & 100\% & -1.0000 \\
    
    S10 & 12.2978 & 66.66\% & 2.1538 & 28.9586 & 56.66\% & 0.9231 & 11.6978 & 100\% & 0.3077 \\
    
    S11 & 10.2205 & 63.33\% & 2.0769 & 11.2106 & 50.00\% & 4.4615 & 9.9977 & 100\% & 3.0769 \\
    
    S12 & 11.2981 & 63.33\% & 1.0000 & 14.6869 & 53.33\% & 0.6154 & 9.9978 & 100\% & -0.2308 \\
    
    S13 & 11.2769 & 90.00\% & -1.8462 & 10.3704 & 90.00\% & -1.6923 & 9.9980 & 100\% & -0.3077 \\
    
    S14 & 11.0100 & 66.66\% & 1.8462 & 17.3993 & 73.33\% & 0.2308 & 9.9977 & 100\% & 1.5385 \\
    
    Average & $12.1 \pm $0.58 & $58.09 \pm $9.69 & $0.98 \pm $0.66 & $15.46 \pm $2.26 & $67.14 \pm $9.39 & $0.54 \pm $0.75 & $11.40 \pm $0.56 & $100 \pm $0 & $0.86 \pm $0.54 \\
	\addlinespace[1ex]
    \hline\hline
  \end{tabular}
  \vspace{1ex}
  \\
  \footnotesize\textsuperscript{a} The displayed SNR values are the combined average SNR values of the different stimuli.\\
  \footnotesize\textsuperscript{b} Fatigue values were baseline corrected based on the initial scores of the participants before the experiment.
\end{table*}

\subsection{VAS-F Questionnaire Analysis Results}
\begin{figure}[!htbp]
  \centering
  \includegraphics[width=.5\linewidth]{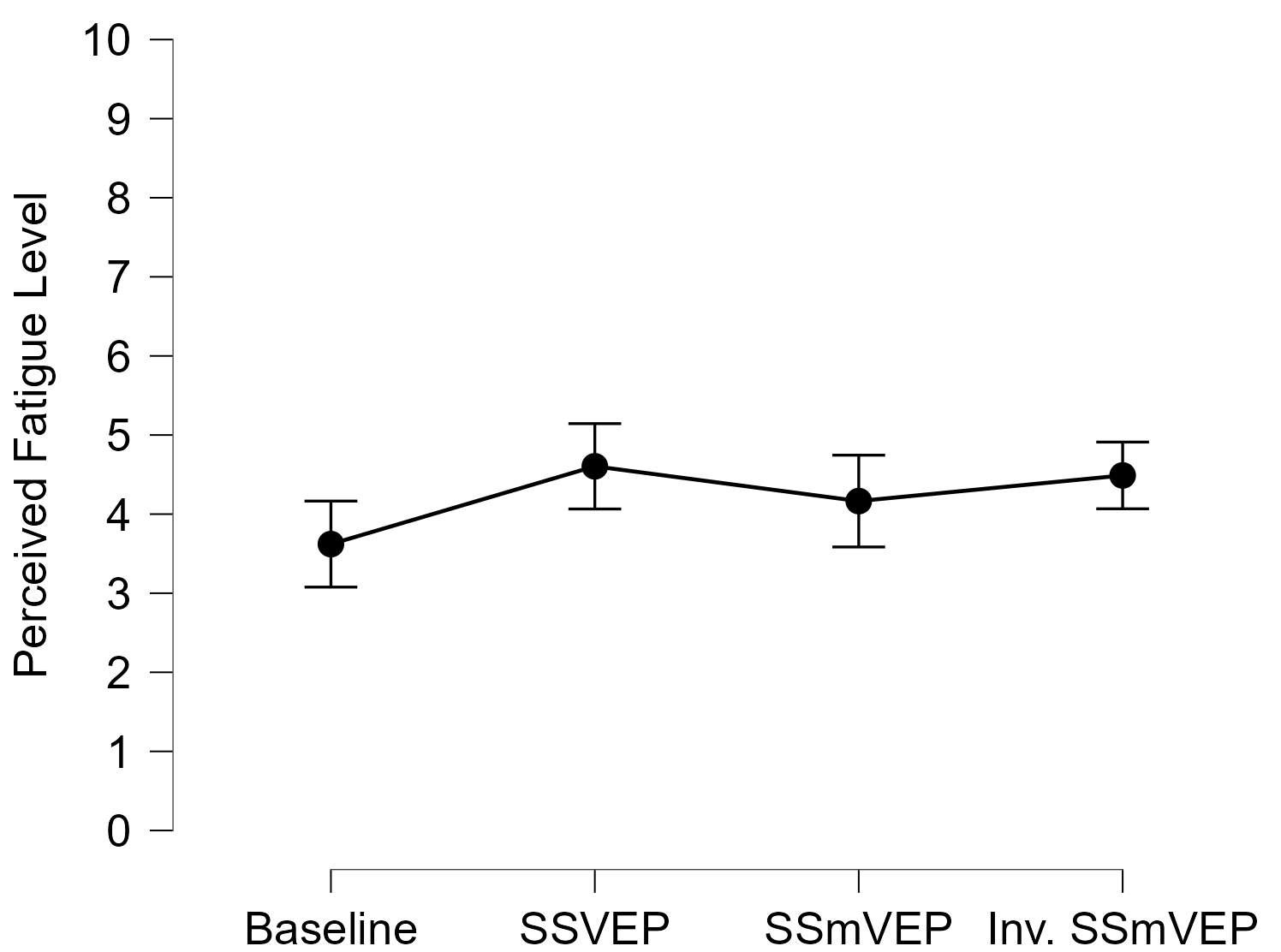}\hfill
  \includegraphics[width=.5\linewidth]{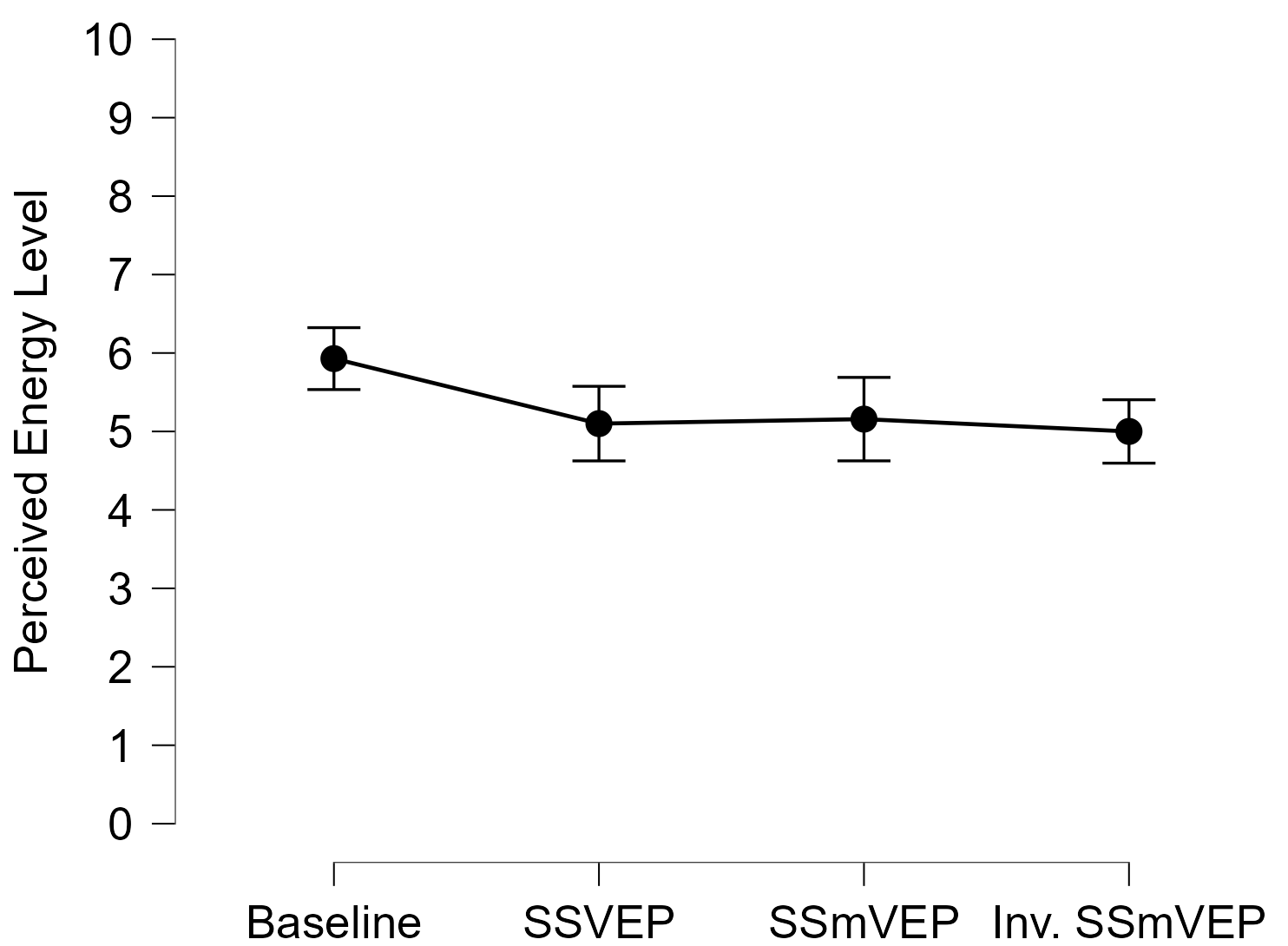}
  \caption{Average fatigue and energy level scores of the participants before the experiment as baseline, after SSVEP trials in Task 1, after SSMVEP trials in Task 2 and after invisible SSMVEP trials in Task 3.}
  \label{fig:vasf}
\end{figure}

In order to assess the fatigue induced by the visible flickering, perceptible motion, and the Gabor-based imperceptible motion stimulation, we employed the Visual Analogue Scale to Evaluate Fatigue Severity (VAS-F) \cite{lee91,shahid11} survey. The survey was administered at the beginning of the experiment session and after each session involving the task with a specific stimuli type. The VAS-F scale includes 18 questions, where 13 questions address fatigue-related indicators, whereas five questions aim to identify how energetic the participants feel. The corresponding items were averaged to obtain fatigue and energy level scores for each participant (Figure \ref{fig:vasf}).

A repeated measures ANOVA analysis on the survey results indicated a significant difference in perceived fatigue levels across rest, flickering, motion, and imperceptible motion stimulation conditions, $F(3, 39)=3.29$, $p<.05$, $\eta^2=.20$. Follow-up post-hoc tests with Holm correction showed that the difference is due to the disparity between baseline fatigue ratings and SSVEP fatigue ratings, $t(13)=-2.87$, $p<.05$, $d=-.49$.

The results also indicated a significant difference in terms of perceived energy level ratings, $F(3, 39)=4.10$, $p<.05$, $\eta^2=.24$. Post-hoc tests suggest that this difference is due to the slight decrease in perceived energy levels during all VEP conditions with respect to the baseline resting condition.

\section{Discussion and Conclusion}
In reactive BCI research, high-frequency flickering or motion-based visual stimuli can lead to fatigue, reduced usability, and, in some cases, trigger photo-sensitive epilepsy. Recent proposals aim to develop imperceptible visual stimuli beyond the Critical Flicker Fusion threshold. In this work, we proposed a sinusoidal grating-based \emph{Gabor patch} stimulus currently utilized only with high-frequency, low-contrast visual properties. However, the parameters that define a Gabor patch, i.e., contrast, angle, phase, spatial frequency, and amplitude, allow for further exploration of how SSMVEP analogous effects in the occipital cortex interact with stimuli dependent features that were previously unavailable to researchers. Additionally, in the presentation of the novel stimuli, we developed a motion animation that is imperceptible when displayed on a high refresh rate monitor in an attempt to evaluate the utilization of SSMVEP in reactive-BCI for future approaches. The results revealed that the Gabor SSMVEP stimulus successfully evoked SSMVEP responses within acceptable margins compared to SSVEP and SSMVEP literature standards. Our statistical analysis results suggest that this novel stimulus type elicits distinguishable peak amplitudes at the targeted frequency range and has comparable SNR levels as compared to perceptible SSVEP and SSMVEP stimuli. Moreover, the observed canonical correlation values show promise for sufficient detection for reactive BCI applications.  

We acknowledge that this study has limitations due to its exploratory nature. A primary constraint is the lack of classification performance analysis among simultaneously presented imperceptible SSMVEP stimuli, which is currently difficult to achieve on a single monitor. In future work, we will develop a suitable setup for presenting multiple imperceptible SSMVEP stimuli to the participants. Furthermore, due to its subjective nature, our fatigue analysis was not sensitive enough to make reasonable comparisons between perceptible and imperceptible VEP conditions. In future studies, we will explore the use of pupillometry-based metrics to improve fatigue analysis.

\section*{Appendix}

Appendices A-C are video files for the three stimulus paradigms. Appendix D visualize the imperceptible motion of the proposed stimulus using high speed cameras.


\begin{thebibliography}{1}
\bibliographystyle{IEEEtran}

\bibitem{adrian34} E. Adrian and D. Matthews, "The Berger rhythm: Potential changes from the occipital lobes in man," \emph{Brain}, vol. 57, no. 3, pp. 355--385, 1934.

\bibitem{regan89} R. Regan, "Human brain electrophysiology: Evoked potentials and evoked magnetic fields in science and medicine," \emph{Elsevier}, 1989.

\bibitem{xie12} J. Xie, G. Xu, J. Wang, F. Zhang, Y. Zhang, "Steady-state motion visual evoked potentials produced by oscillating Newton's rings: implications for brain-computer interfaces," \emph{PLoS ONE}, vol. 7, no. 6, e39707, Jun. 2012.

\bibitem{snowden04} R.J. Snowden and T.C.A. Freeman, "The visual perception of motion," \emph{Curr. Biol.}, vol. 14, no. 19, pp. R828--R831, Oct. 2004.

\bibitem{härdle14} W. K. Härdle and Z. Hlávka, \emph{Canonical Correlation Analysis}, \emph{J. Financial Econ. Policy}, vol. 6, no. 2, pp. 179–196, Apr. 2014.

\bibitem{punsawad12} Y. Punsawad and Y. Wongsawat, "Motion visual stimulus for SSVEP-based BCI system," 2012, pp. 3837--40.

\bibitem{chai19} X. Chai et al., "A radial zoom motion-based paradigm for steady state motion visual evoked potentials," \emph{Front. Hum. Neurosci.}, vol. 13, 2019, p. 127.

\bibitem{jiang22} L. Jiang, W. Pei, and Y. Wang, "A user-friendly SSVEP-based BCI using imperceptible phase-coded flickers at 60Hz," \emph{China Commun.}, vol. 19, no. 2, pp. 1-14, Feb. 2022.

\bibitem{pastor03} M. A. Pastor, J. Artieda, et al., "Human cerebral activation during steady-state visual-evoked responses," \emph{Journal of Neuroscience}, vol. 23, no. 37, 2003, pp. 11621--11627.

\bibitem{bakardjian10} H. Bakardjian, T. Tanaka, G. Cichocki, "Optimization of ssvep brain responses with application to eight\textendash command brain\textendash computer interface," \emph{Neuroscience Letters}, vol. 469, no. 1, pp. 34--38, 2010.

\bibitem{hoffman09} U. Hoffmann, E. J. Fimbel, T. Keller, "Brain-computer interface based on high frequency steady-state visual evoked potentials: A feasibility study," \emph{4th International IEEE/EMBS Conference on Neural Engineering}, 2009.

\bibitem{regan77} D. Regan, "Steady-state evoked potentials," \emph{Journal of the Optical Society of America}, vol. 67, no. 11, 1977, pp. 1475--1489.

\bibitem{williams04} P. E. Williams, F. Mechler, et al., "Erratum: Entrainment to video displays in primary visual cortex of macaque and humans," \emph{Journal of Neuroscience}, vol. 24, no. 44, 2004, pp. 8278--8288.

\bibitem{ladouce22} S. Ladouce, L. Darmet, J. J. Torre Tresols, S. Velut, G. Ferraro, and F. Dehais, "Improving user experience of SSVEP BCI through low amplitude depth and high frequency stimuli design," \emph{Sci. Rep.}, vol. 12, no. 1, p. 8865, May 2022.

\bibitem{zhu10} D. Zhu, J. Bieger, et al., "A survey of stimulation methods used in SSVEP-based BCIs," \emph{Computational Intelligence and Neuroscience}, vol. 2010, 2010, pp. 1--12.

\bibitem{chen19} J. Chen, A. Maye, et al., "Simultaneous decoding of eccentricity and direction information for a single-flicker SSVEP BCI," \emph{Electronics}, vol. 8, no. 12, 2019.

\bibitem{fisher05} R. S. Fisher, G. Harding, et al., "Photic-and pattern-induced seizures: a review for the epilepsy foundation of america working group," \emph{Epilepsia}, vol. 46, no. 9, 2005, pp. 1426--1441.

\bibitem{maye17} A. Maye, D. Zhang, et al., "Utilizing retinotopic mapping for a multi-target SSVEP BCI with a single flicker frequency," \emph{IEEE Transactions on Neural Systems and Rehabilitation Engineering}, vol. 25, no. 7, 2017, pp. 1026--1036.

\bibitem{chen14} X. Chen, Z. Chen, et al., "A high-ITR SSVEP-based BCI speller," \emph{Brain-Computer Interfaces}, vol. 1, no. 3-4, 2014, pp. 181--191.

\bibitem{duszyk14} A. Duszyk, M., Bierzyńska, and Z. Radzikowska, "Towards an optimization of stimulus parameters for brain-computer interfaces based on steady-state visual evoked potentials," \emph{PLoS ONE}, vol. 9, no. 11, 2014, e112099.

\bibitem{chai20} X. Chai et al., "Effects of fatigue on steady state motion visual evoked potentials: Optimised stimulus parameters for a zoom motion-based brain-computer interface", \emph{Computer Methods and Programs in Biomedicine}, vol. 196, p. 105650, 2020.

\bibitem{yan17} W. Yan, G. Xu, et al., "Steady-state motion visual evoked potential (SSMVEP) based on equal luminance colored enhancement," \emph{PLoS ONE}, vol. 12, no. 1, 2017, e0169642.

\bibitem{yan18} W. Yan, G. Xu, et al., "Four novel motion paradigms based on steady-state motion visual evoked potential," \emph{IEEE Transactions on Biomedical Engineering}, vol. 65, no. 8, 2018, pp. 1696--1704.

\bibitem{norcia16} A. M. Norcia, L. G. Applebaum, et al., "The steady-state visual evoked potential in vision research: A review," \emph{Journal of Vision}, vol. 15, no. 6, 2016, p. 4.

\bibitem{ne23} Neuroelectrics, "Neuroelectrics Enobio," \emph{Neuroelectrics} \url{https://www.neuroelectrics.com/solutions/enobio}, Accessed: November 27, 2023.

\bibitem{shahid11} A. Shahid, K. Wilkonson, et al., "Visual analogue scale to evaluate fatigue severity (VAS-F)," \emph{STOP, THAT and one hundred other sleep scales}: Springer, 2011, p. 399--402.

\bibitem{sr23} SR Research Ltd., "Eyelink 1000 Plus," \emph{SR Research}, \url{https://www.sr-research.com/eyelink-1000-plus/}, Accessed: November 27, 2023.

\bibitem{psychopy} J. W. Peirce, J. R. Gray, S. Simpson, M. R. MacAskill, R. H\"{o}chenberger, H. Sogo, E. Kastman, and J. Lindel\o v, "PsychoPy2: Experiments in Behavior Made Easy," \emph{Behavior Research Methods}, vol. 51, no. 1, pp. 195-203, 2019.

\bibitem{odom04} J. V. Odom, M. Bach, C. Barber, et al., ``Visual Evoked Potentials Standard (2004),'' \textit{Documenta Ophthalmologica}, vol. 108, no. 2, pp. 115–123, 2004.

\bibitem{eyelink_sdk} SR Research Ltd., "EyeLink Developers Kit/API," \emph{SR Research} \url{https://www.sr-research.com/software-integration/}, Accessed: November 27, 2023.

\bibitem{nic2} Neuroelectrics, "NIC2," \emph{Neuroelectrics} \url{https://www.neuroelectrics.com/solution/software-integrations/nic2}, Accessed: November 27, 2023.

\bibitem{lsl} Lab Streaming Layer (LSL), \url{https://labstreaminglayer.readthedocs.io/index.html}, Accessed: November 27, 2023.

\bibitem{gramfort13} A. Gramfort, M. Luessi, et al., "MEG and EEG data analysis with MNE-Python," \emph{Frontiers in Neuroscience}, vol. 7, no. 267, 2013, p. 1--13.

\bibitem{cheveigné19} A. de Cheveigné, "ZapLine: A Simple and Effective Method to Remove Power Line Artifacts", \emph{Preprint}, https://doi.org/10.1101/782029, 2019.

\bibitem{mullen15} T. R. Mullen, C. A. E. Kothe, Y. M. Chi, A. Ojeda, T. Kerth, S. Makeig, et al., "Real-time Neuroimaging and Cognitive Monitoring Using Wearable Dry EEG", \emph{IEEE Trans. Bio-Med. Eng.}, vol. 62, pp. 2553–2567, 2015.

\bibitem{bach99}
M. Bach and T. Meigen, ``Do's and don'ts in Fourier analysis of steady-state potentials,'' \textit{Doc Ophthalmol}, vol. 99, pp. 69–82, 1999. DOI: \url{10.1023/A:1002648202420}.

\bibitem{meigen99}
T. Meigen and M. Bach, ``On the statistical significance of electrophysiological steady-state responses,'' \textit{Doc Ophthalmol}, vol. 98, pp. 207–232, 1999. DOI: \url{10.1023/A:1002097208337}.

\bibitem{chen15}
X. Chen, Y. Wang, S. Gao, T. P. Jung, and X. Gao, ``Filter bank canonical correlation analysis for implementing a high-speed SSVEP-based brain-computer interface,'' \textit{Journal of Neural Engineering}, vol. 12, no. 4, pp. 046008, 2015. DOI: \url{10.1088/1741-2560/12/4/046008}.

\bibitem{jasp}
JASP Team, \emph{JASP (Version 0.18.1) [Computer software]}, 2023. \emph{Available}: \url{https://jasp-stats.org/}

\bibitem{barb10}
K. Barb et al., "Welch method revisited: Nonparametric power spectrum estimation via circular overlap," \emph{IEEE Trans. Signal Process.}, vol. 58, no. 2, pp. 553–565, Feb. 2010. [Online].

\bibitem{jukiewicz18}
M. Jukiewicz, M. Buchwald, and A. Cysewska-Sobusiak, "Finding optimal frequency and spatial filters accompanying blind signal separation of EEG data for SSVEP-based BCI," \emph{International Journal of Electronics and Telecommunications}, vol. 64, no. 4, 2018.

\bibitem{lee91}
K. A. Lee, G. Hicks, and G. Nino-Murcia, "Validity and reliability of a scale to assess fatigue," \emph{Psychiatry Research}, vol. 36, no. 3, pp. 291-298, 1991.

\end{thebibliography}
\end{document}